# Short length Lyot filter utilized in dual-wavelength and wavelength tunable mode-locked fiber laser generation

Yuanjun Zhu, Xiangnan Sun, Lei Jin, Shinji Yamashita and Sze Yun Set*

*The University of Tokyo, Research Center for Advanced Science and Technology, 4-6-1 Komaba, Meguro-ku, Tokyo 153-8904, Japan*

E-mail: set@cntp.t.u-tokyo.ac.jp

Dual-wavelength mode-locked fiber laser and wavelength tunable mode-locked fiber laser have emerged as a promising light source. However, there is few reports on generating these two output from just one laser cavity. In this report, we demonstrate a dual-wavelength and wavelength tunable mode-locked fiber laser output from one laser cavity by utilizing short length Lyot filter. The central wavelengths of dual-wavelength mode-locked fiber laser are 1540 nm and 1564 nm and the tunable laser range is about 32 nm. We believe it could provide a simple set-up for generating dual-wavelength and wavelength tunable mode-locked output from one laser cavity.

Dual-wavelength mode-locked fiber laser output via one laser cavity is now attracting more and more attention for the applications in dual-comb laser source, microwave generation and Terahertz wave generation [1-4]. However, because of mode competition between different wavelengths of the gain spectrum, dual-wavelength mode-locked fiber laser is hard to be generated from laser cavity. In order to obtain dual-wavelength mode-locked fiber laser, comb filters including sagnac loop filters, fiber Bragg gratings, Fabry-Perot filters etc. have been inserted into the laser cavity to suppress the mode competition of erbium-doped fiber (EDF) [5-7]. Commonly, the free spectrum range (FSR) of comb filters is too narrow to obtain broad spacing dual-wavelength mode-locked fiber laser. In addition, X. Zhao et al. inserted a programmable attenuator into laser cavity to give gain-tilt to balance two peaks of EDF's gain spectrum [8]. However, by just setting programmable attenuator to be a constant loss is hard to generate stable dual-wavelength mode-locked fiber laser.

On the other hand, wavelength tunable mode-locked fiber lasers have been applied to spectroscopy, optical communications, fiber-optics sensor and biomedical research [9-11]. There are various techniques to achieve wavelength tunable mode-locked laser such as commercial band-pass filter (BPF), fiber Bragg grating etc [12-13]. However, the BPF has a limited bandwidth and large insertion loss. In addition, the fiber Bragg grating is difficult to fabricate and unstable. Recently, Lu et al. proposed an idea to insert 45° tilted fiber grating (45°-TFG) with strong polarization dependent loss to achieve wavelength tunable mode-locked fiber laser [14]. However, the insertion loss of the 45°-TFG is quite large, which makes it hard to generate dual-wavelength mode-locked fiber laser output in the same cavity. The Lyot filter which was invented by Bernard Lyot in 1933 is a kind of polarization interference filter [15]. It can be used as comb filters because of the continuous wavelength-dependent transmission band in the spectra. The Lyot filters have found many applications in spectral imaging [16-17], communication [18] and laser systems [19-22]. Fiber based Lyot filter is commonly composed by a section of polarization maintaining fiber (PMF), a polarization controller (PC) and a polarizer [23-24].

In this report, by choosing a length of 23 cm-long PMF to compose a short length Lyot filter, dual-wavelength and wavelength tunable mode-locked fiber laser are obtained in one laser cavity by adjusting PC. The central wavelengths of dual-wavelength are 1540 and 1564 nm. The range of wavelength tunable mode-locked fiber laser is from 1532 to 1564 nm, which covers the gain spectrum of EDF.

By using Jones matrices, the transmission of Lyot filter is calculated to be written as [25]:

$$T = \frac{1}{2}cos^2(\frac{\pi \Delta n}{\lambda}L_{PMF})(1+sin2\theta) \qquad (1)$$

In this formula, Δn and L_PMF are the birefringence and length of PMF, respectively. θ is the angle between the polarization direction of the input light and the fast axis of the PMF, which can be tuned by adjusting the PC. Therefore, the FSR of Lyot filter can be expressed as:

$$\Delta\lambda = \lambda^2/(\Delta n * L_{PMF}) \qquad (2)$$

The birefringence of PMF is around $3.47 \times 10^{-4}$. From equation (2), the FSR of Lyot filter is inversely proportional to the product of the birefringence and length of PMF. In order to form a broad FSR Lyot filter, two solutions can be proposed. One is using low-birefringence fiber such as bending long length single-mode fiber (SMF) which is reported by our group [26]. The other way is reducing the length of PMF in Lyot filter. Compared with using bending SMF, using short length PMF can make the total cavity length shorter, which is beneficial for high repetition rate laser output.

Here, for obtaining around 30 nm spacing FSR Lyot filter, 23-cm long PMF is chosen. In this way, the short-length Lyot filter is composed of 23-cm long PMF, a fiber-loop type PC and a polarizer. The calculated transmission properties of short length Lyot filter are shown in Fig. 1 (a) and (b) by changing the polarization angles and total birefringence, respectively. In Fig. 1(a), different colors refer to the different polarization angles of PC while keeping the total birefringence constant. This shows that by adjusting the PC, different transmission curves can be achieved, which indicates that the short length Lyot filter can be used to give gain-tilt to balance the two peaks of EDF's gain. In a certain condition, broad spacing dual-wavelength mode-locked fiber laser can be generated. In Fig. 1(b), different colors refer to different total birefringence while setting the polarization angles unchanged. Because the fiber inside the fiber-loop type PC is twisted and squeezed, the birefringence of it can be changed slightly while adjusting the PC [27]. It is shown that the transmission peak's positions can be tuned by slightly changing the total cavity birefringence. In practice, by rotating the PC, both polarization angle and total birefringence will be changed, which indicates the transmission should be the combination of Fig. 1 (a) and (b) with peaks and amplitude tuning function.

In order to observe the transmission property of Lyot filter, a test experiment is employed as follow. A commercial erbium-doped fiber amplifier (EDFA) (AEDFA-PM-23-B-FA) is used as input source and a 50/50 optical coupler (OC) is utilized to split the source into two ways: one goes directly to optical spectrum analyzer (OSA) as reference while the other pass through the short length Lyot filter first and then go to OSA as signal. By using signal power to divide reference power, transmission property of short-length Lyot filter can be observed. The detected transmission of short-length Lyot filter is shown in Fig. 1(c). Here, different colors refer to different states of PC. The traces show different peak's positions and amplitudes, which indicates that this device can introduce gain-tilt to equalize the two peaks of EDF's gain spectrum and tune the central wavelengths of mode-locked laser.

The schematic of the experimental setup is shown in Fig. 2. A 5 m-long erbium-doped silica fiber (EDF) is used as the gain media with an estimated anomalous dispersion of ~-50 ps2/m at 1550 nm. The EDF is pumped through a 980/1550 wavelength-division multiplexing (WDM) coupler in a forward-directional pumping scheme. An isolator (ISO) is used to ensure a unidirectional operation. A carbon nanotube (CNT) solution which is sprayed on the fiber end with the loss of 0.71 dB is inserted into the laser cavity as saturable absorber material. The short length Lyot filter which is composed by a 23-cm long PMF, a fiber-loop-type PC and a polarizer is utilized as wavelength selector to achieve dual-wavelength and wavelength tunable mode-locked operation. A 20/80 OC is used as the output coupler. A 13 m-long length of additional single-mode fiber (SMF) with an estimated anomalous dispersion of ~17 ps2/m at 1550 nm are inserted between the WDM and the OC for dispersion management. The pigtails of all the passive fiber components are made of SMF. The total cavity length is approximately 30 m.

The optical spectrum, temporal waveform, RF spectrum and the autocorrelation trace of the output are measured by an OSA (YOKOGAWA 6375), an oscilloscope (RIGOL DS2202E), a RF spectrum analyzer (RIGOL DSA832) with a 5 GHz photo-detector (PD, DET08CFC/M) and an autocorrelator (Femtochrome FR-103 XL) with a fiber-pigtailed input, respectively.

It's well-known that the EDF's gain spectrum can hold two peaks under certain condition. However, because of the mode competition between the two peaks, always only single-wavelength pulse can be obtained. Here, by inserting the short length Lyot filter whose transmission can be tuned by adjusting the PC into the cavity, the mode competition can be suppressed. The optical spectrum of dual-wavelength mode-locked fiber laser centers at 1540 and 1564 nm is shown in Fig. 3(a). There is one continuous wave (CW) component at 1564 nm pulse, which maybe because the mode-locked condition is not optimized at 1564 nm. The temporal waveform of dual-wavelength mode-locked fiber laser is displayed in Fig. 3(b). The repetition time of it is ~150 ns, which is consistent with the total cavity length of 30 m. The RF spectrum is shown in Fig. 3(c). The repetition rate is 6.38 MHz, which is also consistent with the total cavity length. By zooming in the RF spectrum with the resolution bandwidth (RBW) and video bandwidth (VBW) of 30 Hz, different repetition rates of these two wavelengths pulse can be observed. The difference of them is ~300 Hz, which is due to the group velocity dispersion (GVD) in the cavity. The repetition rate difference can be tuned by changing the GVD in the laser cavity. The signal to noise ratio (SNR) of these two pulses are 42 dB and 43 dB, which indicates that these two pulses have nearly the same intensity.

A tunable optical band-pass filter (BPF) (Alnair Labs BVF-200) which has a flat top band is used to filter out one of the two wavelengths, with the pass band set at 1531 nm to 1549 nm (for 1540nm output) and 1556 nm to 1573 nm (for 1564 nm output), respectively. The optical spectra and autocorrelation traces of the pulses at 1540 nm and 1556 nm can be measured separately. As shown in the Fig. 6(a) and 6(b), the 1532 nm output pulse has a spectral half-width of ~5 nm and an inferred full-width at half-maximum (FWHM) pulse-width of 940 fs assuming a Gaussian waveform. Because the side wings of the pulse spectrum are filtered out by the band-pass filter, the resulting spectrum resemble more like a Gaussian shape rather than the original hyperbolic-sech shape. Also, as shown in the Fig. 6(c) and (d), the spectral half width and the inferred FWHM pulse-width of the 1556 nm pulse component is 4.5 nm and 820 fs, respectively. The time-bandwidth products (TBPs) for the 1532 nm and 1556 nm pulses are 0.504 and 0.457, respectively, which is slightly larger than the transform-limited value of 0.441 assuming Gaussian waveforms. The reason for the non-ideal TBP may be attributed to the the fact that part of the spectral component was filtered out by the BPF. Moreover, because of the high loss induced by the band-pass filter, the measured autocorrelation traces were measured at very low powers impacting the accuracy of the autocorrelation pulse-width measurement.

The wavelength tunable ranging from 1532 to 1564 nm mode-locked fiber laser is shown in Fig. 5. Because of the GVD in the laser cavity, the pulses at different wavelength operate in slightly different Soliton regime. This truly changes the pulse shape of each pulse such as Kelly sidebands. In addition, for the different states of PC, the amplitude of transmission peaks will change, which introduces the different intensity of each pulse. From this figure, 32 nm tunable range which covers the whole gain bandwidth of EDF is observed. Because the Lyot filter has a continuous wavelength-dependent transmission band in the spectra, it is potentially to be applied with other gain medium such as thulium-doped fiber and ytterbium-doped fiber.

In order to confirm the functionality of the short length Lyot filter, the PMF and polarizer are removed from the cavity on purposely. Only single-wavelength mode-locked at 1560 nm can be generated. This truly indicates that the short length Lyot filter is used to balance two peaks of EDF'S gain spectrum and tune the central wavelength of single-wavelength mode-locked laser. In addition, the stability of the dual-wavelength mode-locked fiber laser is tested over 24 hours without any damage.

In summary, we demonstrate a dual-wavelength and wavelength tunable mode-locked fiber laser emitted from one laser cavity using a Lyot filter which is composed of a 23 cm-long PMF, a polarizer and a PC. By adjusting the PC to tune the transmission property of Lyot filter to give gain-tilt to balance the mode competition between two peaks of EDF'S gain spectrum. Dual-wavelength pulse centers at 1540 and 1564 nm can be obtained. In addition, adjusting the PC can change the birefringence of the fiber loop, which can slightly tune the total birefringence of the Lyot filter. Wavelength tunable ranging from 1532 to 1564 nm mode-locked fiber laser can be achieved. We believe it may simplify the set-up for dual-wavelength and wavelength tunable mode-locked fiber from one cavity. It is potentially to be applied to dual-comb generation and fiber-optics sensors.

**Acknowledgments**

This work was supported by JSPS Grant-in-Aid for Scientific Research (S) Grant Number 18H05238.

## References

1) M. Xu, J. Archambault, L. Reekie and J. Dakin, Electronics Letters, 30, 1085-1087 (1994).
2) F. Ganikhanov, S. Carrasco, X. S. Xie, M. Katz, W. Seitz, and D. Kopf, Opt. Lett. 31, 1292-1294 (2006).
3) L. Talaverano, S. Abad, S. Jarabo and M. Lopez-Amo, Journal of Lightwave Technology, 19, 553-558 (2001).
4) Thi Van Anh Tran, Kwanil Lee, Sang Bae Lee, and Young-Geun Han, Opt. Express 16, 1460-1465 (2008).
5) C. Kim, R. M. Sova, and J. U. Kang, Opt. Commun. 218, 291–295 (2003).
6) Z. Wang, Y. Cui, B. Yun, and C. Lu, IEEE Photonics Technol. Lett., 17, 2044–2046 (2005).
7) S. Pan, C. Lou, and Y. Gao, Opt. Express 14, 1113–1118 (2006).
8) X. Zhao, Z. Zheng, L. Liu, Y. Liu, Y. Jiang, X. Yang, and J. Zhu, Opt. Express 19, 1168-1173 (2011).
9) N. Chen, J. Lin, F. Liu and S. Liaw, in IEEE Photonics Technol. Lett., 22, 700-702 (2010).
10) X. Li, Y. Wang, Y. Wang, X. Hu, W. Zhao, X. Liu, J. Yu, C. Gao, W. Zhang, Z. Yang, C.

Li and D. Shen, in IEEE Photonics Journal, 4, 234-241 (2012).

11) H. Zhang, D. Tang, R. Knize, L. Zhao, Q. Bao and K. Loh, Appl. Phys. Lett., 96, 111-112 (2010).

12) D. Li, H. Jussila, Y. Wang, G. Hu, T. Albrow-Owen, R. C. T. Howe, Z. Ren, J. Bai, T. Hasan, and Z. Sun, Sci. Rep. 8, 2738 (2018).

13) X. He, Z. Lin and D. Wang, Opt. Lett., 37, 2394-2396 (2012).

14) B. Lu, C. Zou, Q. Huang, Z. Yan, Z. Xing, M. Al Araimi, A. Rozhin, K. Zhou, L. Zhang and C. Mou, Journal of Lightwave Technology, 37, 3571-3578 (2019).

15) B. Lyot, C. R. Acad. Sci. (Paris) 197, 1593 (1933).

16) A. Gorman and D. Fletcher-Holmes, Opt. Express 18, 5602 (2010).

17) O. Aharon and I. Abdulhalim, Opt. Express 17, 11426 (2009).

18) M. Huang, J. Chen, K. Feng, C. Wei, C. Lai, T. Lin, and S. Chi, IEEE Photon. Technol. Lett. 18, 172 (2006).

19) C. O'Riordan, M. J. Connelly, and Pr. M. Anandarajah, Opt. Commun. 281, 3538 (2008).

20) M. Franke, W. Paa, W. Triebel, and H. Stafast, Appl. Phys. B 97, 421 (2009).

21) K. Özgören and F. Ö. Ilday, Opt. Lett. 35, 1296-1298 (2010).

22) N. Park and P. F. Wysocki, in IEEE Photonics Technology Letters, 8, 1459-1461 (1996).

23) Z. Yan, C. Mou, H. Wang, K. Zhou, Y. Wang, W. Zhao, and L. Zhang, Opt. Lett. 37, 353-355 (2012).

24) S. Sugavanam, Z. Yan, V. Kamynin, A. S. Kurkov, L. Zhang, and D. V. Churkin, Opt. Express 22, 2839-2844 (2014).

25) S. Liu, F. Yan, F. Ting, L. Zhang, Z. Bai, W. Han, and H. Zhou, IEEE Photonics Technol. Lett., 28, 864-867 (2016).

26) Y. Zhu, F. Xiang, L. Jin, S. Y. Set and S. Yamashita, IEEE Photonics Journal, 11(6), 1-7 (2019).

27) W. Man, H. Tam, M. Demokan, P. Wai, and D. Tang, J. Opt. Soc. Am. B 17, 28-33 (2000).

**Figure Captions**

**Fig. 1.** Simulated transmission of short length Lyot filter by changing (a) polarization angles and (b) total birefringence, (c) Experimental test of the transmission of Lyot filter (different colors refer to different PC states).

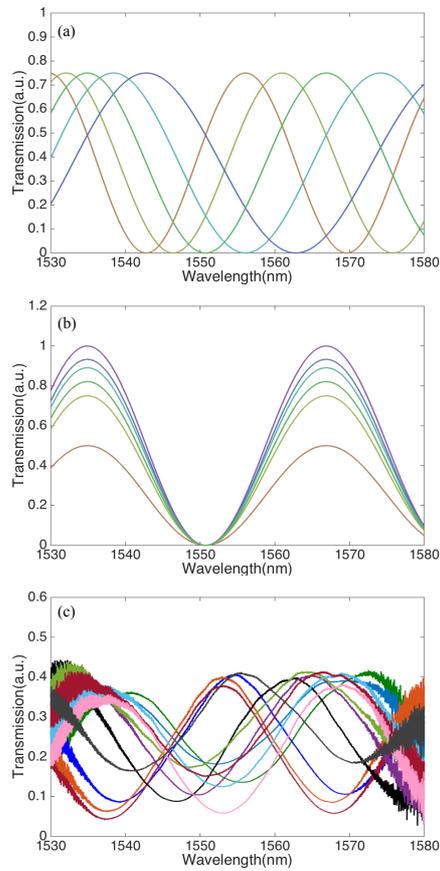

**Fig. 2.** Schematic setup of the ring-cavity laser by inserting CNT-SA and short length Lyot filter. (WDM: wavelength-division multiplexer; EDF: erbium-doped fiber; PI-ISO: polarization-insensitive isolator; PMF: polarization maintaining fiber; PC: polarization controller; OC: optical coupler; SMF: single-mode fiber).

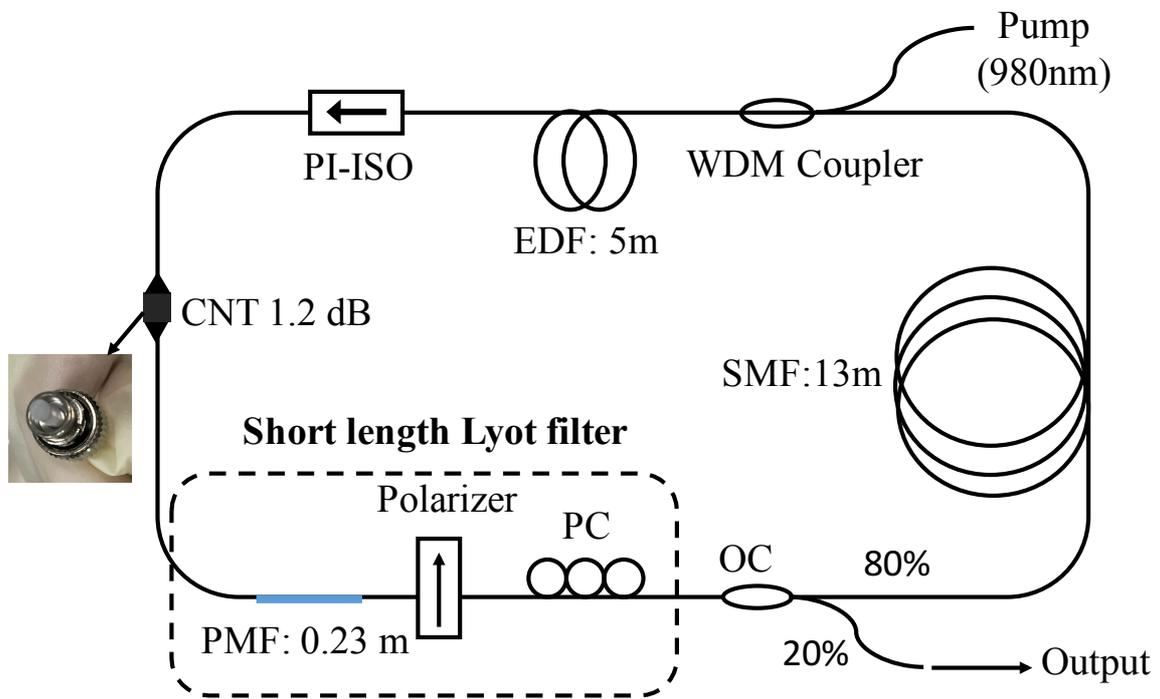

**Fig. 3.** (a) Output spectra (Resolution 0.05 nm) (b) Temporal waveform (c) (RBW=VBW= 1kHz) and (d) (RBW=VBW=30Hz) RF spectrum of dual-wavelength mode-locked fiber laser.

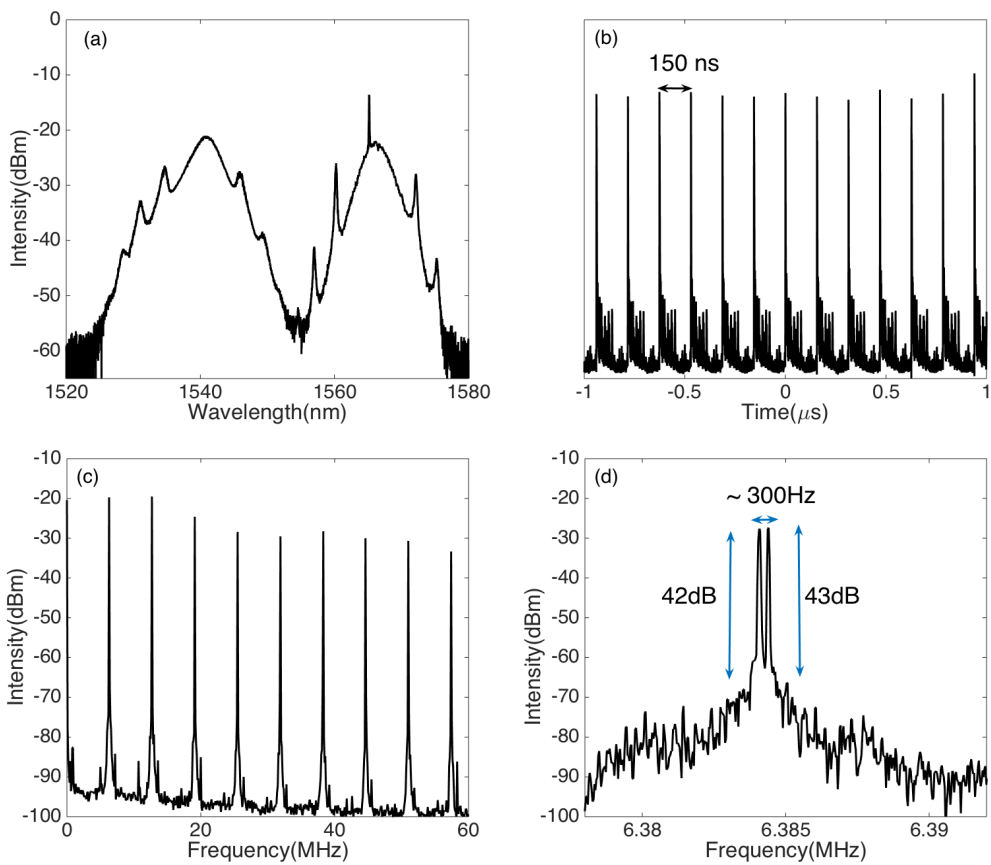

**Fig. 4.** Optical spectrum (Resolution: 0.05 nm) of (a) 1540 nm pulse and (b) 1564 nm pulse. Autocorrelation traces of (c) 1540 nm pulse and (d) 1564 nm pulse.

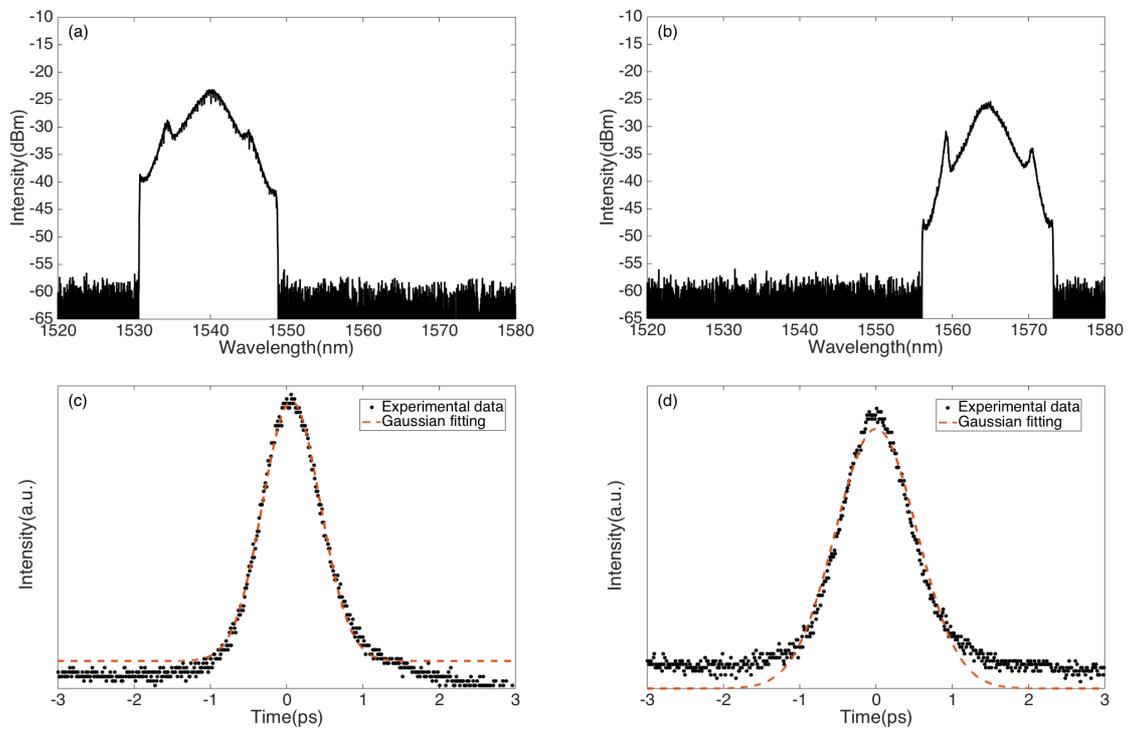

**Fig. 5.** Output spectra of tunable mode-locked laser from 1532 nm to 1564 nm. (Resolution: 0.05 nm).

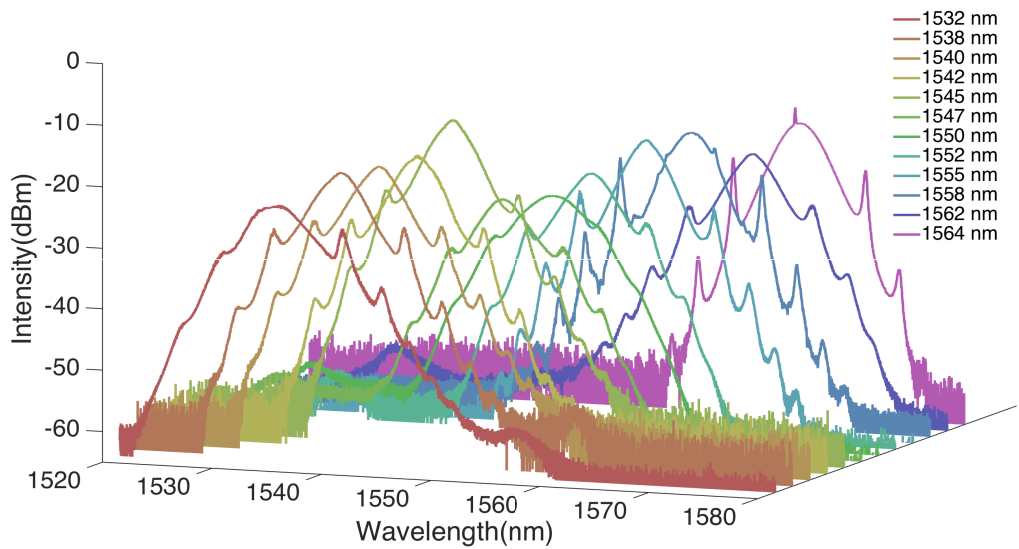